\documentclass[]{spie}  

\usepackage{amsmath,amsfonts,amssymb}
\usepackage{graphicx}
\usepackage[colorlinks=true, allcolors=blue]{hyperref}
\usepackage{tikz}
\usepackage{booktabs}
\usepackage{siunitx}
\usepackage{subcaption}

\title{Optical design of a direct fibre-fed MO-IFS for the NLOT}
\author[a]{Nitish Singh}
\author[a]{S. Sriram}
\author[a]{Totan Chand}
\author[b]{Jurgen Schmoll}
\author[a]{Bharat Kumar Yerra}
\author[a]{Savitha M S}
\affil[a]{Indian Institute of Astrophysics, II Block, Koramangala, Bengaluru 560 034, INDIA}
\affil[b]{Centre for Advanced Instrumentation, Durham University}

\authorinfo{Further author information: Residence: Bhaskara, Indian Institute of Astrophysics Guest House, Koramangala II Block, Bengaluru 560 034, India\\
Nitish Singh: E-mail: nitish.singh@iiap.res.in, nitiphy@gmail.com\\
Telephone: +91 8077759357}

\begin{document} 
\maketitle

\begin{abstract}
The initial optical design and performance analysis of a dual-channel fibre-fed Multi-Object Integral Field spectrograph (Mo-IFS) being designed for a future National Large Optical/Infrared Telescope (NLOT) in India. The front end will be a moveable lenslet+fiber based integral field unit. The spectrograph is designed to directly accept an f/4 beam from the approximately 200 optical fibers, each with a 100 $\mu m$ core diameter without additional fore-optics. The instrument consists of two optimized spectral channels covering wavelength ranges of 0.32–0.62 $\mu m$ (blue channel) and 0.60–1.00 $\mu m$ (red channel). The optical design aims to achieve moderate spectral resolutions of approximately R $\sim$ 2700 in the blue channel and R $\sim$ 2500 in the red channel while maintaining high throughput over a broad spectral range. The spectrograph architecture includes a fiber-fed entrance slit, collimator optics, dichroic beam splitting system, dispersive elements, and dedicated camera optics for each channel. Zemax simulations were carried out to evaluate image quality, spot size distribution, spectral resolution, and detector sampling across the full wavelength range. The current work presents the initial optical configuration, design methodology, and expected performance of the instrument.
\end{abstract}



\section{INTRODUCTION}
\label{sec:intro}  

Spectroscopy plays a fundamental role in modern observational astronomy by providing detailed information about the physical and chemical properties of astronomical objects \cite{1984aste.book.....K, 2013ias..book.....A}. Through spectroscopic observations, it is possible to study stellar atmospheres, galaxy evolution, star formation regions, active galactic nuclei, and the large-scale structure of the Universe \cite{2013ias..book.....A}, \cite{1998gaas.book.....B}. The increasing demand for large-scale spectroscopic surveys and time-efficient observations has driven the development of fiber-fed spectrographs capable of simultaneously observing multiple targets with high stability and flexibility \cite{2019BAAS...51g.274K, 2001MNRAS.328.1039C}. Fiber-fed spectrographs have become an essential component of modern astronomical instrumentation due to several advantages, including improved mechanical stability, flexible instrument placement, efficient light transmission, and simplified integration with telescope focal planes \cite{2005PASP..117.1411F, 2004SPIE.5492..624S}. In particular, direct fiber-fed spectrographs provide compact optical architectures by directly coupling the telescope beam delivered by optical fibers into the spectrograph optics without additional relay optics \cite{2013AJ....146...32S}. Such systems can reduce optical losses and simplify alignment requirements while maintaining good throughput and spectral performance. The next generation of large optical telescopes requires spectrographs capable of covering broad wavelength ranges with moderate spectral resolution for a wide variety of scientific programs. Simultaneous blue and red wavelength coverage is especially important for observing spectral features distributed over large wavelength intervals. For this, we are designing a multi-object integral field spectrograph (Mo-IFS). It's a Dual-channel spectrographs employing dichroic beam splitters provide an efficient solution for achieving wide spectral coverage while maintaining optimized optical performance in individual wavelength bands. In this work, we present the initial optical design and performance analysis of a dual-channel direct fiber-fed spectrograph being developed for a future large telescope in India \cite{2024JApA...45....8P, 2022JApA...43...32A}. The spectrograph architecture includes a fiber-fed entrance slit, collimator optics, a dichroic beam splitter, dispersive elements, and dedicated camera optics for each spectral channel. Optical optimization and performance evaluation were carried out using Zemax optical design software. The present paper discusses the instrument concept, optical design methodology, and expected imaging and spectral performance of the proposed spectrograph.

\section{Instrument Concept and Design Requirements}

\label{sec:current_coupl}

\begin{figure}[htbp]
    \centering
    \includegraphics[width=0.8\linewidth]{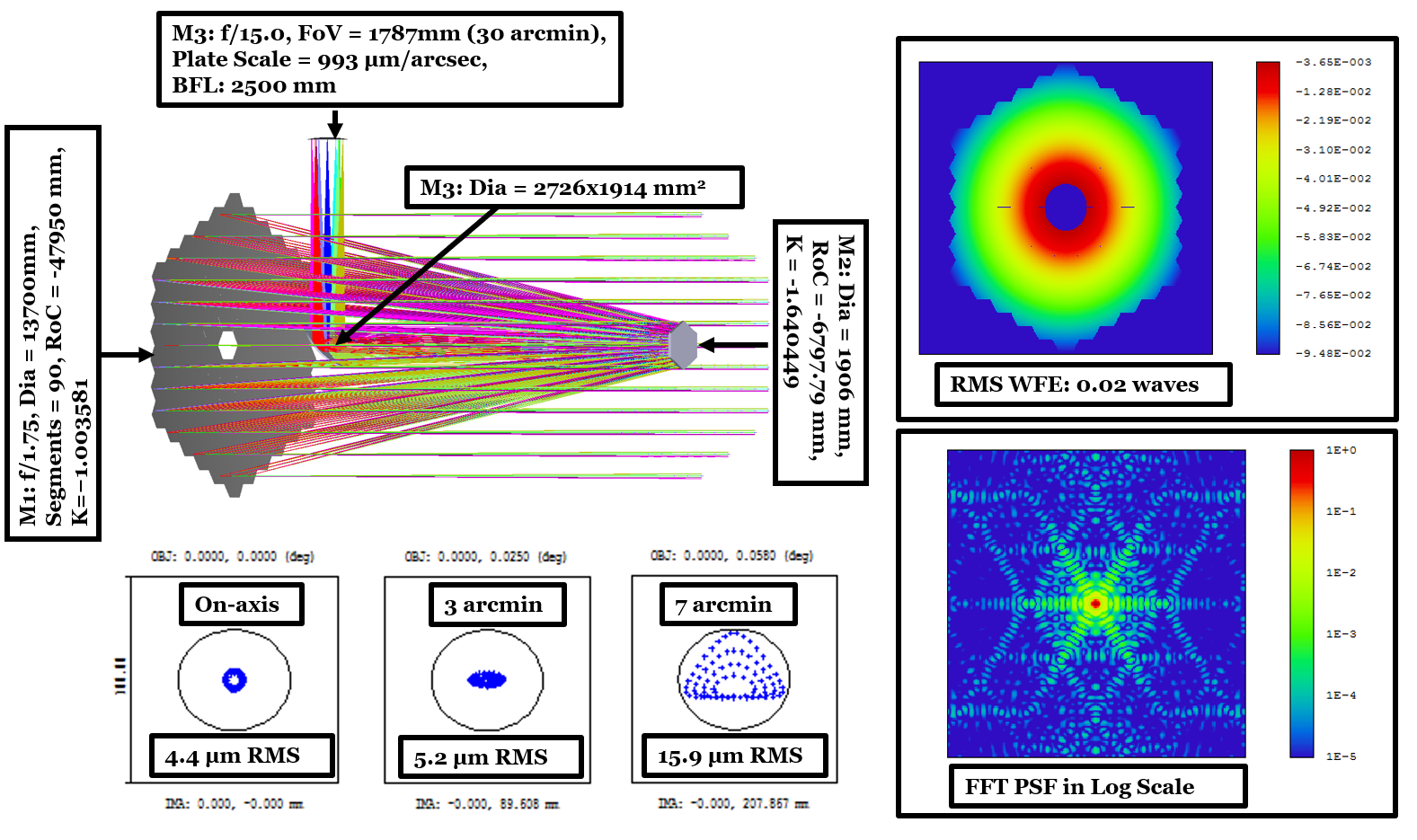}
    \caption{Optical Layout of NLOT}
    \label{fig:NLOT_Optical_Design}

\end{figure}

The National Large Optical/Infrared Telescope (NLOT) is a planned 13.7 m ground-based optical/infrared telescope based on a modified Ritchey–Chrétien design \cite{2024JApA...45....8P}. The system delivers a Nasmyth focus with an f/15 output beam, providing a wide 20 arcmin field of view enabling high-resolution imaging and spectroscopy across the visible and near-infrared bands \footnote{\href{https://www.iiap.res.in/projects/nlot/overview/}{IIA NLOT Overview}}. The optical design is shown in Figure~\ref{fig:NLOT_Optical_Design}.

\begin{figure}[htbp]
    \centering

    \begin{subfigure}{0.48\linewidth}
        \centering
        \includegraphics[width=\linewidth]{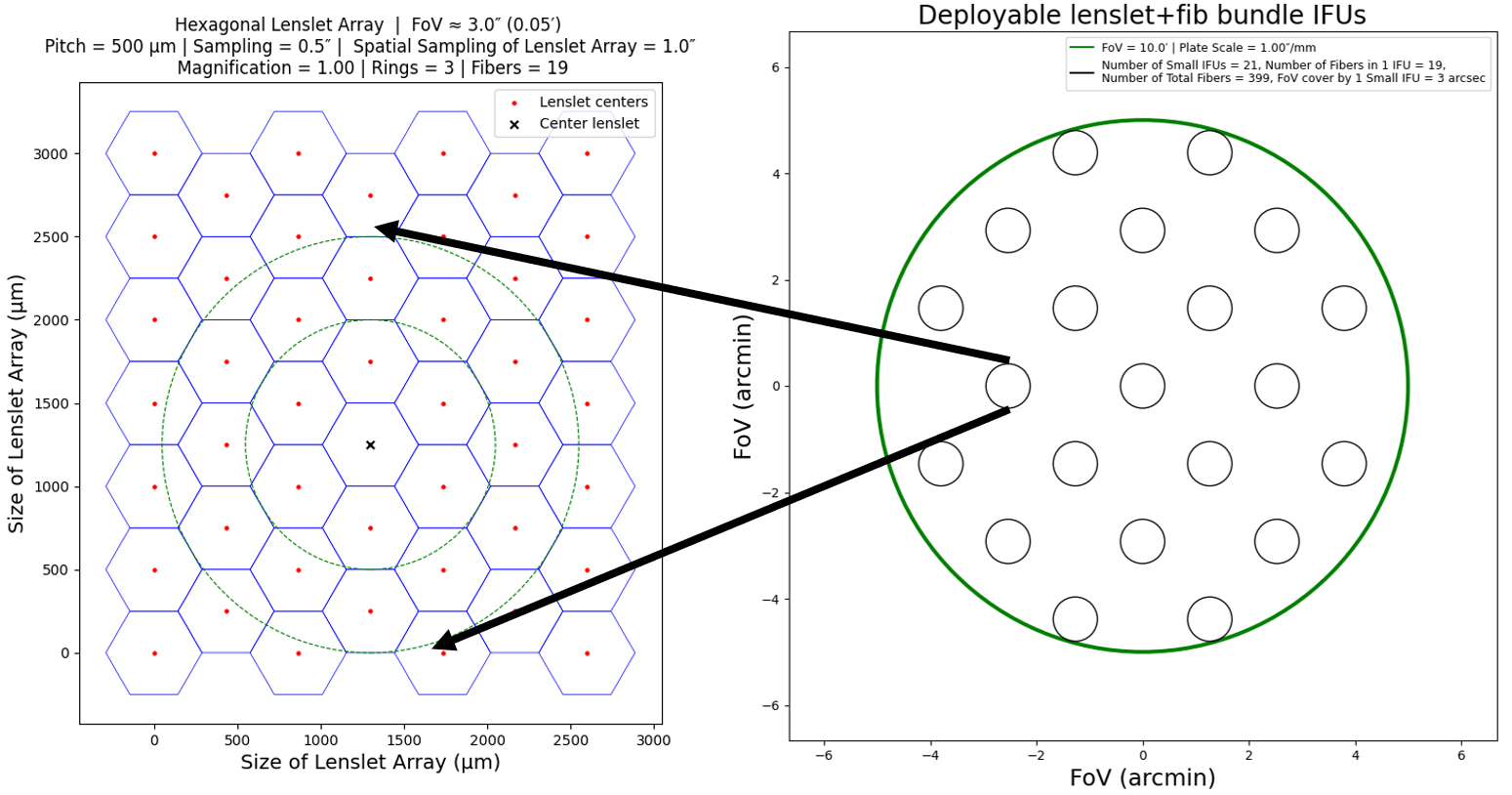}
        \caption{ }
        \label{fig:Deployable_IFU}
    \end{subfigure}
    \hfill
    \begin{subfigure}{0.48\linewidth}
        \centering
        \includegraphics[width=\linewidth]{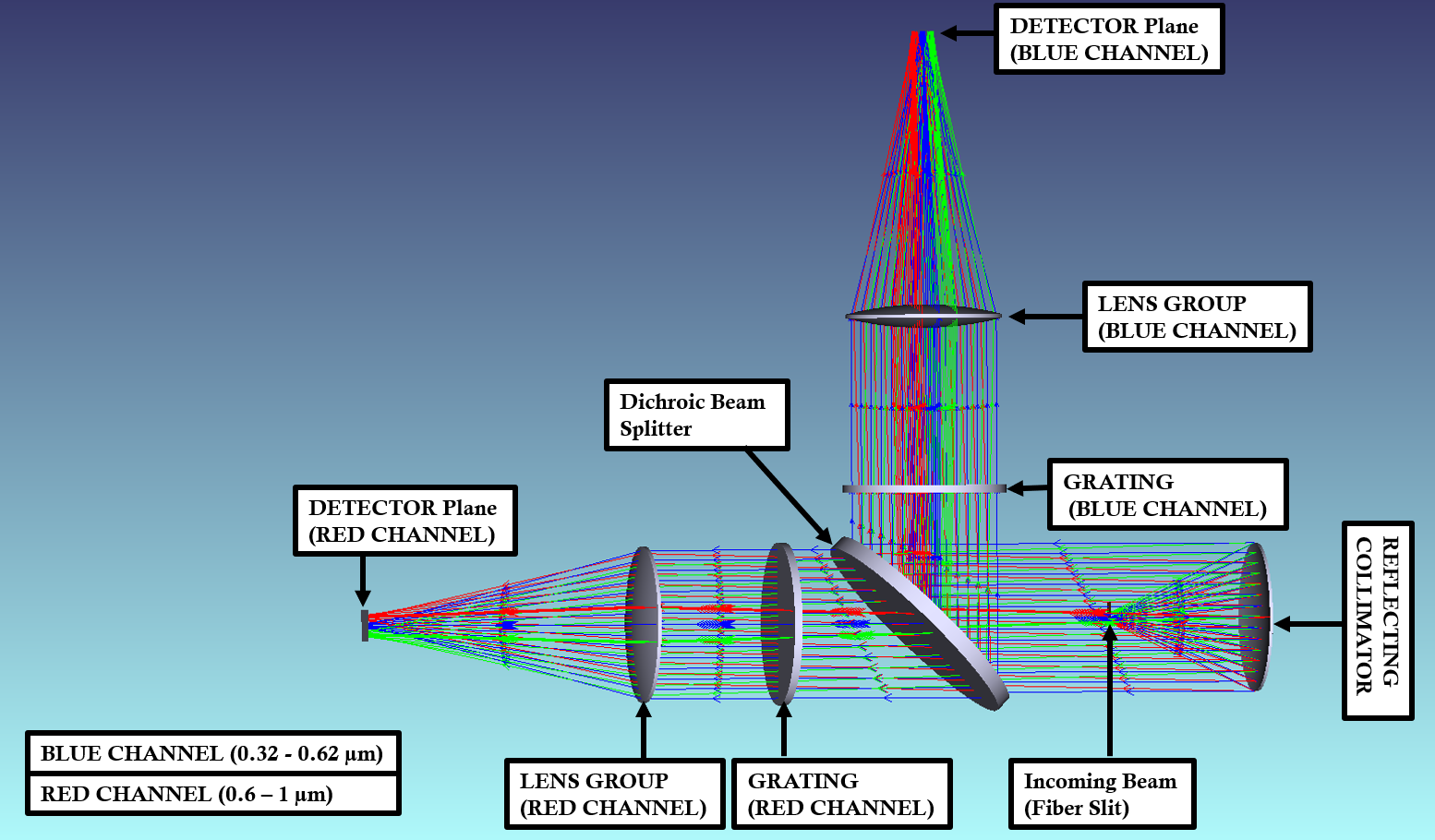}
        \caption{ }
        \label{fig:overview_spectrograph}
    \end{subfigure}

    \caption{(a) Deployable Lenslet+Fiber bassed IFU concept and (b) Overall spectrograph overview optical design for the MO-IFS system.}
    \label{fig:moifs_overview}
\end{figure}

The NLOT wavelength coverage spans 0.31–2.5 $\mu m$. To exploit this capability, we are developing a multi-object integral field spectrograph (Mo-IFS) for the optical regime. The concept uses deployable lenslet-based IFUs coupled to optical fibers, positioned at the Nasmyth focus using robotic fiber positioners. The positioning system can employ either SCARA-type fiber positioners \cite{2020IJO....14...53K,2024SPIE13096E..0IT} or Starbug-type fiber positioners \cite{2010SPIE.7739E..1EG}. The output fibers are routed to a spectrograph located on a stable platform, enabling flexible multi-object integral field spectroscopy across the field of view. A schematic of the deployable IFU system is shown in Figure~\ref{fig:Deployable_IFU}, illustrating the lenslet–fiber coupled IFU at the focal plane, with fibers routed to the spectrograph. The current deployable IFU concept consists of approximately 400 fibers. The complete deployable IFU system is designed to accommodate approximately 400 fibers using two identical 200 fiber spectrograph modules. The present work focuses on the design and performance of a single 200 fiber direct fiber-fed spectrograph. The instrument adopts a direct fiber-fed spectrograph architecture. Unlike traditional slit-based systems, which suffer from flux losses due to slit truncation and variable seeing conditions, the direct fiber approach improves throughput by transmitting the full fiber output to the spectrograph. In typical slit-based designs, these losses can approach 40–50\%, depending on slit width and observing conditions \cite{2024SPIE13100E..4OS, 2024arXiv240615592S}. The proposed configuration avoids this limitation while maintaining stable and efficient coupling from the telescope focal plane to the detector. The spectrograph consists of a fiber pseudo-slit, collimator, diffraction gratings, and camera optics for two spectral channels. A dichroic beam splitter after the collimator separates the beam into blue and red channels, allowing independent optimization of coatings, dispersers, and camera designs as shown in Figure~\ref{fig:overview_spectrograph}. 

The system is designed to handle approximately 200 fibers, each with a 100 $\mu m$ core diameter, accepting an f/4 input beam. The pseudo-slit geometry ensures adequate separation between adjacent spectra on the detector to minimize cross-talk and enable reliable extraction. The optical fiber operate over a broad wavelength range extending from 0.310 $\mu m$ to 1 $\mu m$. The spectrograph consists of two optimized channels covering 0.320–0.620 $\mu m$ in the blue arm and 0.600–1.0 $\mu m$ in the red arm. The optical design targets moderate spectral resolutions of approximately R $\sim$ 2700 and R $\sim$ 2500 for the blue and red channels, respectively.

\section{Optical Parameters Calculation for MO-IFS Design }

The baseline MO-IFS configuration is designed for direct fiber-fed Mo-IFS as shown in Figure~\ref{fig:moifs_overview}. The adopted instrument parameters are summarized in Table~\ref{tab:system_parameters}.

\begin{table}[h]
\centering
\caption{Baseline spectrograph system parameters}
\label{tab:system_parameters}
\begin{tabular}{ll}
\toprule
\textbf{Parameter} & \textbf{Value} \\
\midrule
Collimator diameter ($D_{col}$) & 103 mm \\
Collimator Conic Constant ($e$) & -0.1447 \\
Collimator focal ratio ($F_{\#col}$) & 4 \\
Camera focal ratio ($F_{\#cam}$) & 1.8 \\
Detector pixel size ($p$) & 15 $\mu$m \\
Fiber core diameter ($d_{core}$) & 100 $\mu$m \\
Number of fibers ($N_{fib}$) & 200 \\
Blue channel central wavelength ($\lambda_c$) & 460 nm \\
Blue channel grating density ($G$) & 1200 lines/mm \\
Blue channel groove spacing ($\sigma_{blue}$) & 0.0008333~\text{mm} \\
Red channel central wavelength ($\lambda_c$) & 800 nm \\
Red channel grating density ($G$) & 650 lines/mm \\
Red channel groove spacing ($\sigma_{red}$) & 0.001538~\text{mm} \\
\bottomrule
\end{tabular}
\end{table}

The focal lengths of the collimator ($f_{col}$) and camera optics $f_{cam}$ estimated from the product of the optical diameter and focal ratio, as described in Equation~\ref{eq:focal_length}.

\begin{equation}
f_{col} = D_{col} \cdot F_{\#col} = 103 \times 4 \approx 412~\text{mm},
\qquad
f_{cam} = D_{col} \cdot F_{\#cam} = 103 \times 1.8 \approx 185.4~\text{mm}
\label{eq:focal_length}
\end{equation}

The optical magnification between the fiber plane and detector plane is obtained from the ratio of the collimator and camera focal lengths as described in Equation~\ref{eq:magnification}.

\begin{equation}
M = \frac{f_{col}}{f_{cam}} = \frac{412}{185.4} \approx 2.22
\label{eq:magnification}
\end{equation}

Using this magnification, the projected fiber image size on the detector is estimated as in Equation~\ref{eq:image_size}.

\begin{equation}
\Delta x = \frac{d_{core}}{M} = \frac{0.1}{2.22} \approx 0.045~\text{mm} = 45~\mu m
\label{eq:image_size}
\end{equation}

The projected fiber image size corresponds to the detector sampling in pixels. For a detector pixel size of $15~\mu$m, the sampling is calculated as shown in Equation~\ref{eq:sampling}.

\begin{equation}
\Delta x_{pixel} = \frac{45}{15} \approx 3~\text{pixels}
\label{eq:sampling}
\end{equation}

Thus, the fiber image is adequately sampled over three pixels, ensuring stable spectral extraction and sufficient sampling of the spectral resolution element as shown in Equation~\ref{eq:sampling}.

The diffraction angle at the central wavelength of each channel, where $\lambda_{cblue} = 0.460~\mu m$ and $\lambda_{cred} = 0.80~\mu m$, is estimated using the grating equation, as shown in Equation~\ref{eq:grating_equation}.

\begin{equation}
\beta = \arcsin\left(\frac{m \lambda_c}{\sigma}\right)
\qquad
\beta_{blue} = 33.5^\circ
\qquad
\beta_{red} = 31.33^\circ
\label{eq:grating_equation}
\end{equation}

Using the diffraction angle and groove spacing, the linear spectral dispersion at the detector plane is calculated using Equation~\ref{eq:linear_dispersion}.

\begin{equation}
\frac{d\lambda}{dx} = \frac{\sigma \cos\beta}{m f_{cam}},
\qquad
\left(\frac{d\lambda}{dx}\right)_{blue} = 0.56~\text{\AA/pixel},
\qquad
\left(\frac{d\lambda}{dx}\right)_{red} = 1.07~\text{\AA/pixel}
\label{eq:linear_dispersion}
\end{equation}

where $m$ is the diffraction order and $\sigma$ is the groove spacing of the grating. The spectral resolution element ($\delta \lambda$) corresponding to the projected fiber image size is then estimated using Equation~\ref{eq:spectral_dispersion}.

\begin{equation}
\Delta \lambda = \frac{\sigma \cos(\beta)}{m f_{cam}} \Delta x,
\qquad
\Delta \lambda_{blue} = 1.68~\text{\AA},
\qquad
\Delta \lambda_{red} = 3.18~\text{\AA}
\label{eq:spectral_dispersion}
\end{equation}

Using Equation~\ref{eq:spectral_dispersion}, we calculate $\delta \lambda$, and the spectral resolving power at the central wavelength of each channel is then obtained using Equation~\ref{eq:resolving_power}.

\begin{equation}
R = \frac{\lambda}{\Delta \lambda}
\qquad
R_{blue} = \frac{4600}{1.68} \approx 2738,
\qquad
R_{red} = \frac{8000}{3.18} \approx 2515
\label{eq:resolving_power}
\end{equation}

\subsection{Fiber Separation Constraint}

To minimize spectral overlap and cross-talk between adjacent fibers, the fiber spacing at the pseudo-slit is defined using a separation criterion corresponding to approximately $3\sigma$ of the projected fiber image profile. Assuming a Gaussian profile, the relation between the full width at half maximum (FWHM) and the standard deviation was then calculated using the relation given in Equation~\ref{eq:FWHM_deviation}\cite{2019sto..book.....T}.

\begin{equation}
\mathrm{FWHM} = 2.355\sigma
\label{eq:FWHM_deviation}
\end{equation}

Here, the FWHM corresponds to the fiber core diameter ($d_{core}$) which is $100~\mu$m, and we maintain a $3\sigma$ separation between adjacent fibers, similar to the SDSS/BOSS fiber-fed spectrograph \cite{2013AJ....146...32S}. The separation between adjacent fibers is then calculated using Equation~\ref{eq:specing_fiber}.

\begin{equation}
s = \frac{3 d_{core}}{2.355} \approx 127~\mu m
\label{eq:specing_fiber}
\end{equation}

The total pseudo-slit height for 200 fibers is then estimated using Equation~\ref{eq:slit_height}.

\begin{equation}
h = N_{fib} \cdot s \approx 26~\text{mm}
\label{eq:slit_height}
\end{equation}

This slit height defines the physical extent of the pseudo-slit assembly and provides an important mechanical constraint for the spectrograph design. The corresponding angular slit height is estimated using the small-angle approximation, as shown in Equation~\ref{eq:slit_height_anguler}.

\begin{equation}
\theta = \arctan\left(\frac{h}{f_{col}}\right)
= \arctan\left(\frac{26}{413}\right)
\approx 3.6^\circ
\label{eq:slit_height_anguler}
\end{equation}

These first-order calculations establish the basic geometrical and spectral limits of the spectrograph design and provide the starting point for the detailed optical optimization that follows.

\section{Optical Design}
\label{sec:new_desi}

\begin{figure}[htbp]
        \centering
        \includegraphics[width=0.9\linewidth]{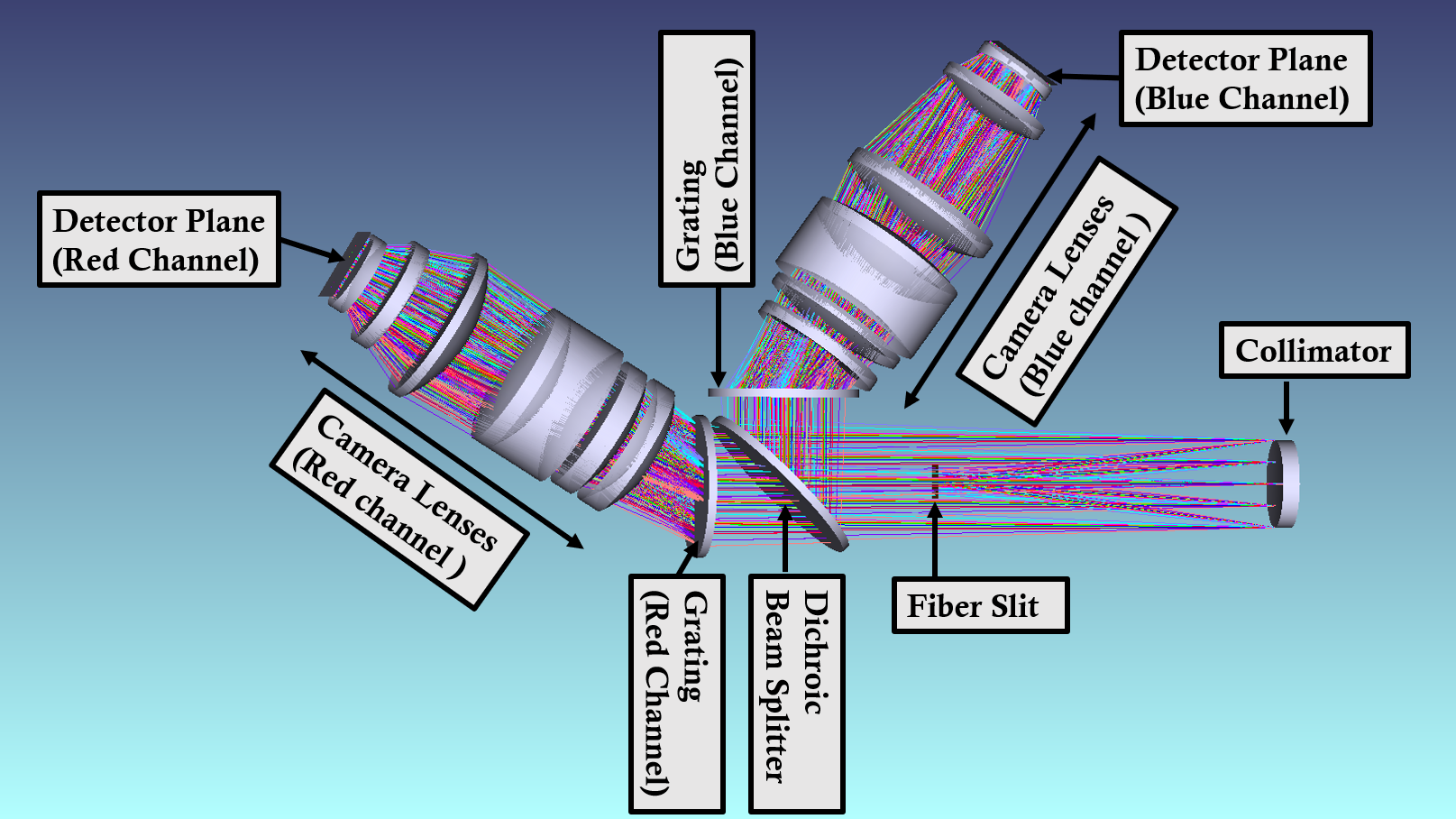}
    \caption{MO-IFS optical layout designs for blue and red channels.}
    \label{fig:moifs_layout}
\end{figure}

The optical design of the spectrograph was developed using the system parameters listed in Table~\ref{tab:system_parameters}. The instrument adopts a direct fiber-fed dual-channel configuration optimized for medium-resolution spectroscopy over the 0.32--1.0 $\mu$m wavelength range. The optical layout begins with a pseudo-slit formed by the output ends of the fibers. The diverging beam emerging from the fiber pseudo-slit is collimated using an ellipsoidal reflective collimator, producing a collimated beam suitable for spectral dispersion. A dichroic beam splitter placed after the collimator separates the beam into blue and red spectral channels. The blue channel covers the 0.32--0.62 $\mu$m range, while the red channel operates over 0.60--1.0 $\mu$m. The optical layouts of both channels are shown in Fig.~\ref{fig:moifs_layout}.

Each channel consists of an independent diffraction grating and refractive camera optics optimized for its respective wavelength range. The use of separate optical paths enables independent optimization of coatings, dispersers, and camera performance for the blue and red bands. The dispersed light from each channel is finally focused onto dedicated detectors with 15 $\mu$m pixel size.

\section{Performance Analysis}

The optical performance of the spectrograph was evaluated for both the blue and red channels over the full pseudo-slit height of approximately 25 mm, corresponding to a field extent of about $\pm 1.95^\circ$ (i.e., from $-1.95^\circ$ to $+1.95^\circ$). The performance analysis was carried out to verify image quality, spectral resolution, and spot containment across the entire slit extent.

\begin{figure}[htbp]
    \centering

    \begin{subfigure}{0.48\linewidth}
        \centering
        \includegraphics[width=\linewidth]{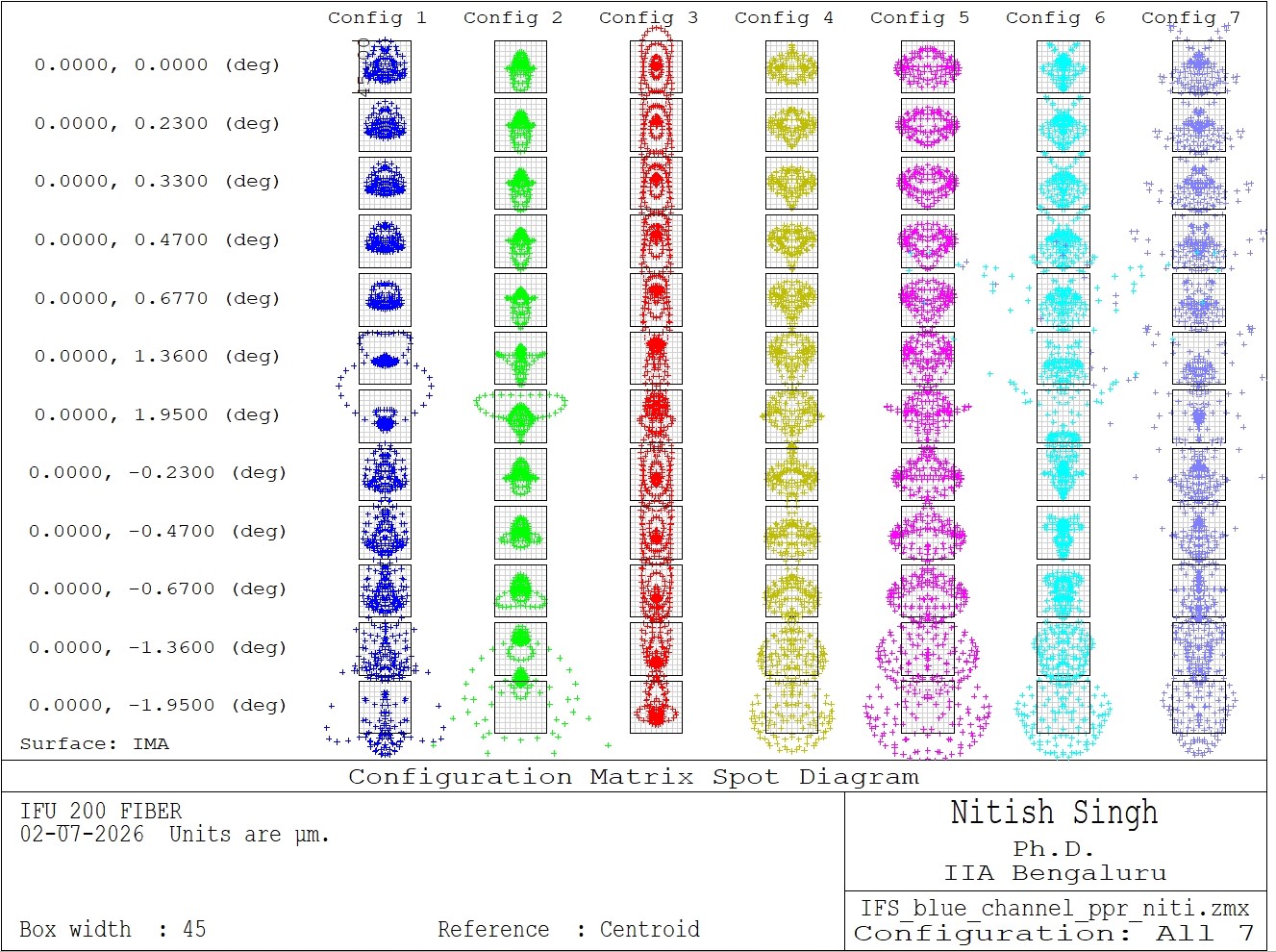}
        \caption{Blue channel configuration matrix}
        \label{fig:blue_channel_performance}
    \end{subfigure}
    \hfill
    \begin{subfigure}{0.48\linewidth}
        \centering
        \includegraphics[width=\linewidth]{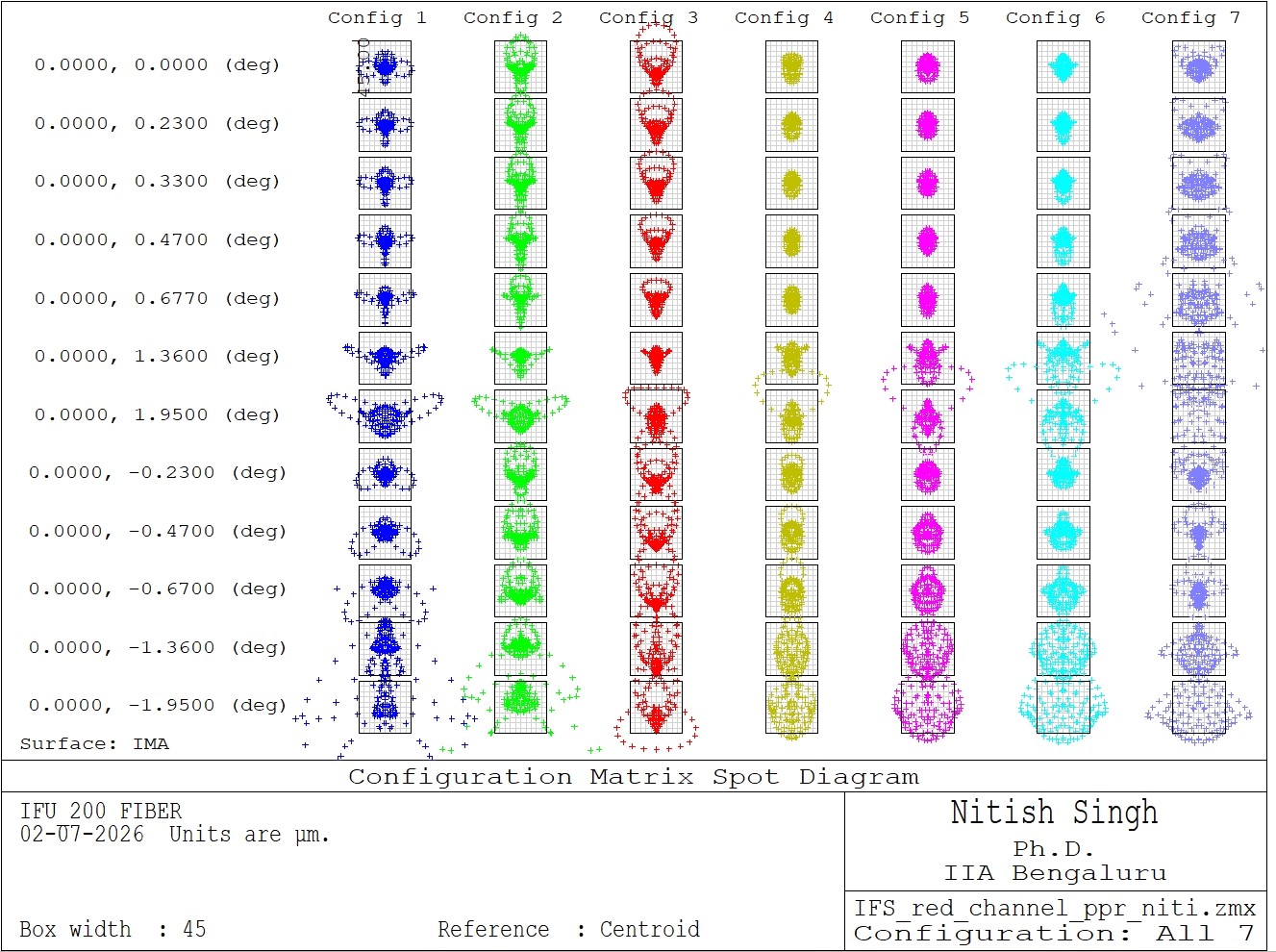}
        \caption{Red channel configuration matrix}
        \label{fig:red_channel_performance}
    \end{subfigure}

    \caption{MO-IFS performance showing Box size 45$\mu m$ (a) blue channel configuration matrices  (b) red channel configuration matrices.}
    \label{fig:moifs_performance}
\end{figure}

For the blue channel, the performance was analyzed over the 0.32--0.62 $\mu$m wavelength range using a 1200 lines/mm diffraction grating. Similarly, the red channel performance was evaluated over the 0.60--1.0 $\mu$m wavelength range using a 650 lines/mm grating configuration. The performance is shown in the configuration matrices for both channels in Fig.~\ref{fig:moifs_performance}.

In the blue channel configuration matrix, each column represents a different wavelength configuration at 0.32, 0.35, 0.40, 0.46, 0.50, 0.60, and 0.62~$\mu$m. In the red channel configuration matrix, each column represents a different wavelength configuration at 0.60, 0.62, 0.70, 0.80, 0.85, 0.90, and 1.00~$\mu$m. The rows correspond to different field positions across the pseudo-slit, covering the full field range from $-1.95^\circ$ to $+1.95^\circ$ for both channel shown in Figure~\ref{fig:moifs_performance}. 

The configuration matrix spot diagrams show the projected fiber image size on the detector, corresponding to approximately 45 $\mu$m or 3 pixels. The design goal was to ensure that the majority of the spectral energy remains confined within the projected fiber footprint across the full slit height. Before presenting the ZEMAX-based results, we first ensure that the theoretical linear dispersion obtained from Equation~\ref{eq:linear_dispersion} is consistent with the linear separation obtained from the ZEMAX model. For this purpose, the theoretical spectral resolution element for both channels is derived using Equation~\ref{eq:spectral_dispersion}. Two closely spaced wavelengths around the central wavelength are then selected for each channel, and the corresponding spectral separation is determined as described in Equation~\ref{eq:spectral_dispersion}. These wavelength pairs are subsequently used in the ZEMAX model to compute the resulting linear separation on the detector. Ideally, this separation should be approximately 45~$\mu$m, corresponding to the projected fiber image size..

\subsection{Linear Dispersion Analysis using ZEMAX: Blue Channel}

\begin{figure}[htbp]
    \centering
    \includegraphics[width=0.8\linewidth]{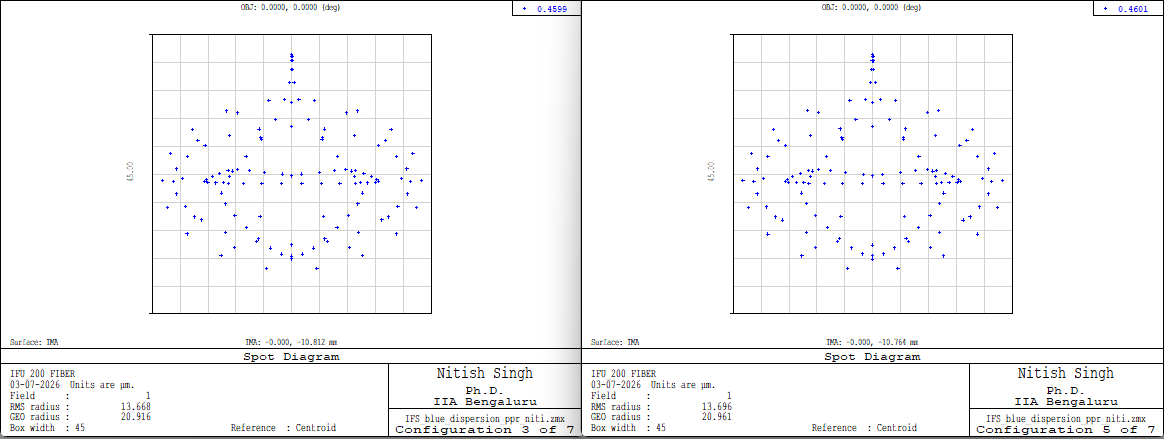}
    \caption{Spot positions on the detector corresponding to $\lambda_{1,blue} = 0.45916~\mu m$ and $\lambda_{2,blue} = 0.46084~\mu m$ for center fiber.}
    \label{fig:blue_spot}
\end{figure}

The spectral performance of the blue channel was evaluated in ZEMAX using two nearby wavelengths around the spectrograph channel's central wavelength, $\lambda_c = 0.460~\mu m$. The corresponding spectral resolution element is 1.68~\AA. The detector positions corresponding to the selected wavelengths are shown below in Equation~\ref{eq:blue_lambda1},\ref{eq:blue_lambda2} and illustrated in Figure~\ref{fig:blue_spot}.

\begin{equation}
\lambda_{1,blue} = 0.45916~\mu m,
\qquad
x_{1,blue} = -11.030~\text{mm}
\label{eq:blue_lambda1}
\end{equation}

\begin{equation}
\lambda_{2,blue} = 0.46084~\mu m,
\qquad
x_{2,blue} = -10.547~\text{mm}
\label{eq:blue_lambda2}
\end{equation}

The spectral separation on the detector is therefore calculated as described in Equation~\ref{eq:X_blue_definition}

\begin{equation}
X_{blue} = x_{1,blue} - x_{2,blue} = 0.048~\text{mm} = 48~\mu m
\label{eq:X_blue_definition}
\end{equation}

\subsection{ZEMAX Performance Calculation: Red Channel}

\begin{figure}[htbp]
    \centering
    \includegraphics[width=0.8\linewidth]{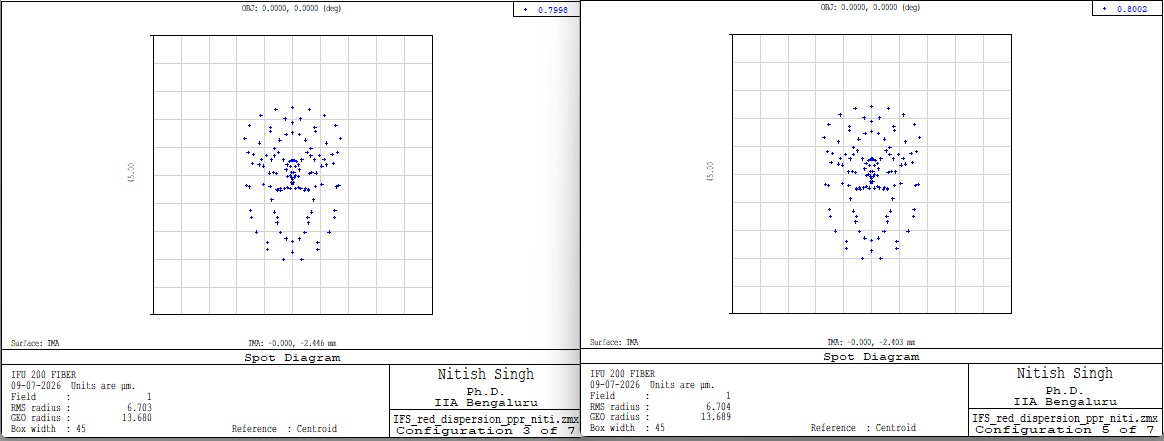}
    \caption{Spot Size and positions on the detector corresponding to $\lambda_{1,red} = 0.799841~\mu m$ and $\lambda_{2,red} = 0.800159~\mu m$ for center fiber}
    \label{fig:red_spot}
\end{figure}

The spectral performance of the red channel was evaluated in ZEMAX using two nearby wavelengths around the spectrograph channel's central wavelength, $\lambda_c = 0.800~\mu m$. The corresponding spectral resolution element is 3.18~\AA. The detector positions corresponding to the selected wavelengths are shown below in Equation~\ref{eq:red_lambda1},\ref{eq:red_lambda2} and illustrated in Figure~\ref{fig:red_spot}.

\begin{equation}
\lambda_{1,red} = 0.799841~\mu m,
\qquad
x_{1,red} = -2.446~\text{mm}
\label{eq:red_lambda1}
\end{equation}

\begin{equation}
\lambda_{2,red} = 0.800159~\mu m,
\qquad
x_{2,red} = -2.403~\text{mm}
\label{eq:red_lambda2}
\end{equation}

The spectral separation on the detector is therefore calculated as described in Equation~\ref{eq:X_red_definition}

\begin{equation}
X_{red} = x_{1,red} - x_{2,red} = 0.043~\text{mm} = 43~\mu m
\label{eq:X_red_definition}
\end{equation}

From Equation~\ref{eq:X_blue_definition}, the spatial separation between two adjacent fiber spectra in the blue channel is approximately $48~\mu$m, corresponding to about 3.2 pixels. Similarly, from Equation~\ref{eq:X_red_definition}, the spatial separation in the red channel is approximately $43~\mu$m, corresponding to about 2.86 pixels. Since the red channel achieves a spatial separation of less than 3 pixels, the blue channel is currently being further optimized to reduce its spatial separation to below 3 pixels as well.
 
\section{Conclusion}

A dual-channel fiber-fed MO-IFS concept has been developed and analyzed for the optical wavelength range of 0.32--1.0~$\mu$m. The design employs a direct fiber-fed pseudo-slit, an ellipsoidal reflective collimator, and a dichroic beam-splitting configuration to enable simultaneous blue and red channel operation. The optical performance was evaluated using ZEMAX ray-tracing, and the results were compared with theoretical estimates of linear dispersion derived from grating-based formulations. The analysis shows good agreement between the theoretical linear dispersion and the ZEMAX-derived linear separation at the detector for both channels. The resulting spectral sampling is consistent with the projected fiber image size of approximately 45~$\mu$m, corresponding to nearly 3 detector pixels. The current ZEMAX analysis provides detector separations of approximately 48~$\mu$m for the blue channel and 43~$\mu$m for the red channel. Further optical optimization is ongoing to reduce the spot size below 45~$\mu$m, corresponding to less than 3-pixel sampling across both channels. The spot distributions remain well confined within the expected fiber footprint across the full pseudo-slit height, confirming good image quality and spectral stability. In parallel, development of the front-end deployable IFU system is also in progress, where movable IFUs will be positioned at the telescope focal plane using robotic fiber positioners. The present design is based on a 200-fiber configuration; however, future development aims to scale the system to nearly 500 fibers to enable higher multiplexing capability for wide-field multi-object integral field spectroscopy.

\section{Declarations}

\subsection{Availability of Data and Materials} \label{material_availibility}  

The data supporting the findings of this study are available from the corresponding author, Nitish Singh, upon reasonable request.

\subsection{Funding declaration}
This research was made possible through financial support from the Tata Consultancy Services Research Fellowship and Indian Institute of Astrophysics (IIA) under the Department of Science and Technology (DST) in India.

\subsection{Declaration of competing interest}

The authors declare no conflict of interest with any known people.

\bibliography{report} 
\bibliographystyle{spiebib} 

\end{document}